\documentclass[conference,keeplastbox, onecolumn]{IEEEtran}
\usepackage{cite}
\usepackage{amsmath,amssymb,amsfonts}
\usepackage{algorithmic}
\usepackage{graphicx}
\usepackage{textcomp}
\usepackage{xcolor}
\usepackage{balance}
\usepackage{multicol}
\usepackage[colorlinks=true,linkcolor=black,anchorcolor=black,citecolor=black,filecolor=black,menucolor=black,runcolor=black,urlcolor=black]{hyperref}
\def\BibTeX{{\rm B\kern-.05em{\sc i\kern-.025em b}\kern-.08em
    T\kern-.1667em\lower.7ex\hbox{E}\kern-.125emX}}
\begin{document}

\title{Building a Sustainable Structure for Research Software Engineering Activities}

\author{\IEEEauthorblockN{Jeremy Cohen\IEEEauthorrefmark{1}, Daniel S.~Katz\IEEEauthorrefmark{2}, Michelle Barker\IEEEauthorrefmark{3}, Robert Haines\IEEEauthorrefmark{4} and Neil Chue Hong\IEEEauthorrefmark{5}} \IEEEauthorblockA{\IEEEauthorrefmark{1}Department of Computing, Imperial College London, London, UK\\
Email: jeremy.cohen@imperial.ac.uk} \IEEEauthorblockA{\IEEEauthorrefmark{2}NCSA, CS, ECE, iSchool,
University of Illinois Urbana-Champaign, 
Urbana, IL, USA \\
Email: d.katz@ieee.org}
\IEEEauthorblockA{\IEEEauthorrefmark{3}Australian Research Data Commons, Cairns, Australia \\
Email: michelle.barker@unimelb.edu.au} \IEEEauthorblockA{\IEEEauthorrefmark{4}Research IT, 
University of Manchester, 
Manchester, UK \\
Email: robert.haines@manchester.ac.uk}
\IEEEauthorblockA{\IEEEauthorrefmark{5}Software Sustainability Institute, EPCC,
University of Edinburgh,
Edinburgh, UK\\
Email: N.ChueHong@software.ac.uk}
}

\maketitle

\textbf{A shortened two-page version of this paper was published by and \copyright 2018 IEEE, as part of the 9th International Workshop on Sustainable Software for Science: Practice and Experiences (WSSSPE6.1) -- \url{http://wssspe.researchcomputing.org.uk/wssspe6-1/}, available at \url{https://doi.org/10.1109/eScience.2018.00015}.}

\begin{multicols}{2}

% 135 words
\begin{abstract}
The profile of research software engineering has been greatly enhanced by developments at institutions around the world to form groups and communities that can support effective, sustainable development of research software. We observe, however, that there is still a long way to go to build a clear understanding about what approaches provide the best support for research software developers in different contexts, and how such understanding can be used to suggest more formal structures, models or frameworks that can help to further support the growth of research software engineering. This paper sets out some preliminary thoughts and proposes an initial high-level model based on discussions between the authors around the concept of a set of pillars representing key activities and processes that form the core structure of a successful research software engineering offering. 
\end{abstract}


\section{Introduction}

While researchers, including academic faculty and staff, postdocs, students, and those working in industry, have been building software to support their research for many decades, their primary goal is generally their research outputs, not the software. There has, however, been significant growth in the number of individuals who are interested in the development of the software itself and the process of working with researchers to help design and build quality, sustainable software. This has led to the fairly recent emergence of the concept of Research Software Engineering. Developed out of discussions that took place at the UK Software Sustainability Institute's~\cite{ref:ssi} Collaborations Workshop in 2012~\cite{ref:ssi-cw12}~\cite{ref:rse-dr2012}, the concept builds on the fact that developing research software requires increasingly advanced skill-sets that must be built up over time by individuals who specialise in the process of writing code and the application of best practices to ensure its reliability and sustainability. Jim\'{e}nez et al.~\cite{ref:Jimenez17} provide an example of four such best practices. While researchers can still teach themselves to code and build up a base of knowledge that enables them, for example, to start analysing their research data or developing user interfaces to support their users, advances in computing hardware and infrastructure, and vast increases in data volumes, raise a number of significant challenges in building research software. While the capabilities of modern computing hardware and new models of computation, such as remote cloud computing infrastructure or GPUs and FPGAs, present significant opportunities to researchers, they also present significant technical barriers.

To take advantage of improvements in technology, and the speed of change in the field, developers need a much more advanced set of knowledge, which takes longer to build and maintain, in order to ensure that they can support research requirements. This is especially true in the case of developers working independently. Software teams sharing knowledge across a group of developers may offer a more manageable way to sustain expertise and to better support specialisation in particular areas or techniques. The additional technical complexity of larger projects and the time required to gain the necessary technical expertise mean that developing some code alongside one's research is becoming increasingly difficult to do well for all but the smallest projects. This has led to a new class of individuals, Research Software Engineers (RSEs), who generally have a research background but have chosen to focus on the software development-related aspects of research. In addition to their knowledge of the research lifecyle, RSEs apply professional software engineering practices in a manner suited to the research environment, following best practices with a view to developing better quality, more sustainable and maintainable research software. The discussions that led to the concept of RSEs emerging observed the special nature of the roles that these individuals hold, but also their challenges in trying to find approaches for career structures and career progression that could make these roles sustainable~\cite{ref:rse-history}.

Ensuring that these structures develop and that there is sustainability for RSEs is still very much work-in-progress. Nonetheless, the profile of research software engineering has been greatly enhanced by the activities occurring at institutions around the world to develop groups and communities that can support more effective, sustainable development of research software. This process was initiated in the UK with the establishment of the first research software engineering groups in various academic institutions, providing a central team of RSEs to undertake software development work for researchers within their local institution. The approach of building institutional research software engineering groups can offer a team structure and scope for career progression, something that is much more challenging for the lone ``researcher-developer'' who is based within, or leading, a research group. These developments have been followed by the emergence of formal initiatives in the UK to champion and advocate for research software engineering: the UK RSE Association~\cite{ref:uk-rse} in 2013, the first EPSRC RSE Fellowships in 2015 and the first RSE Conference in 2016. This has led to the global rise of research software engineering activities including  the first international  workshop for leaders, from across the world, of such research software engineering  groups and communities (e.g., NL-RSE~\cite{ref:intro-nlrse} in the Netherlands, de-RSE~\cite{ref:de-rse} in Germany and RSE-AU~\cite{ref:rse-au} in Australia) which took place in London in 2018~\cite{ref:int-rse-leaders}. This was aimed at people running (or setting up) RSE groups and communities around the world, with participants from Europe, North and South America, Africa and Australasia attending to share experiences and start collaborations. Representatives of the Moore-Sloane Data Science Environments, which are involved in the establishment of new research career structures for RSEs in the USA~\cite{ref:rse4datasci}, also participated.

The UK Research Software Engineer Network (RSEN)'s 2017 State of the Nation Report~\cite{ref:rse-state-of-the-nation} provides a background to the development of research software engineering, as well as a range of statistics about the RSE role and community. To gain a better understanding about RSEs and answers to questions such as what they do, how they do it and how they view their role, a number of surveys of RSEs have now been carried out in various countries~\cite{ref:rse-surveys-data}, for example, in Germany~\cite{ref:de-rse-survey}, Australia and New Zealand~\cite{ref:au-nz-rse-survey}, and the US~\cite{ref:urssi-rse-survey}.

As research software engineering has grown as a concept, it has become clear that there are a number of activities that are common between different offerings at different institutions. It is also clear to us that research software engineering is, or at least should be, about a lot more than individuals writing research software. In this paper we set out the basis for a model that we are currently developing that defines a set of ``pillars'' which encapsulate the core activities that we feel are crucial in ensuring sustainable, long-term support for effective development of software for research. This paper is intended to stimulate further discussion around this area and to support the development of a further publication detailing the next iteration of work defining a complete multi-pillar model. In Section~\ref{section:rs-activities} we highlight the core activities that we see as underlying comprehensive research software engineering support while Section~\ref{section:initial-model} shows how these are brought together in our initial sustainable RSE framework model. Section~\ref{section:conclusions} presents initial conclusions and suggests future work.

\begin{figure*}[hbt]
\centering
\includegraphics[width=0.65\linewidth]{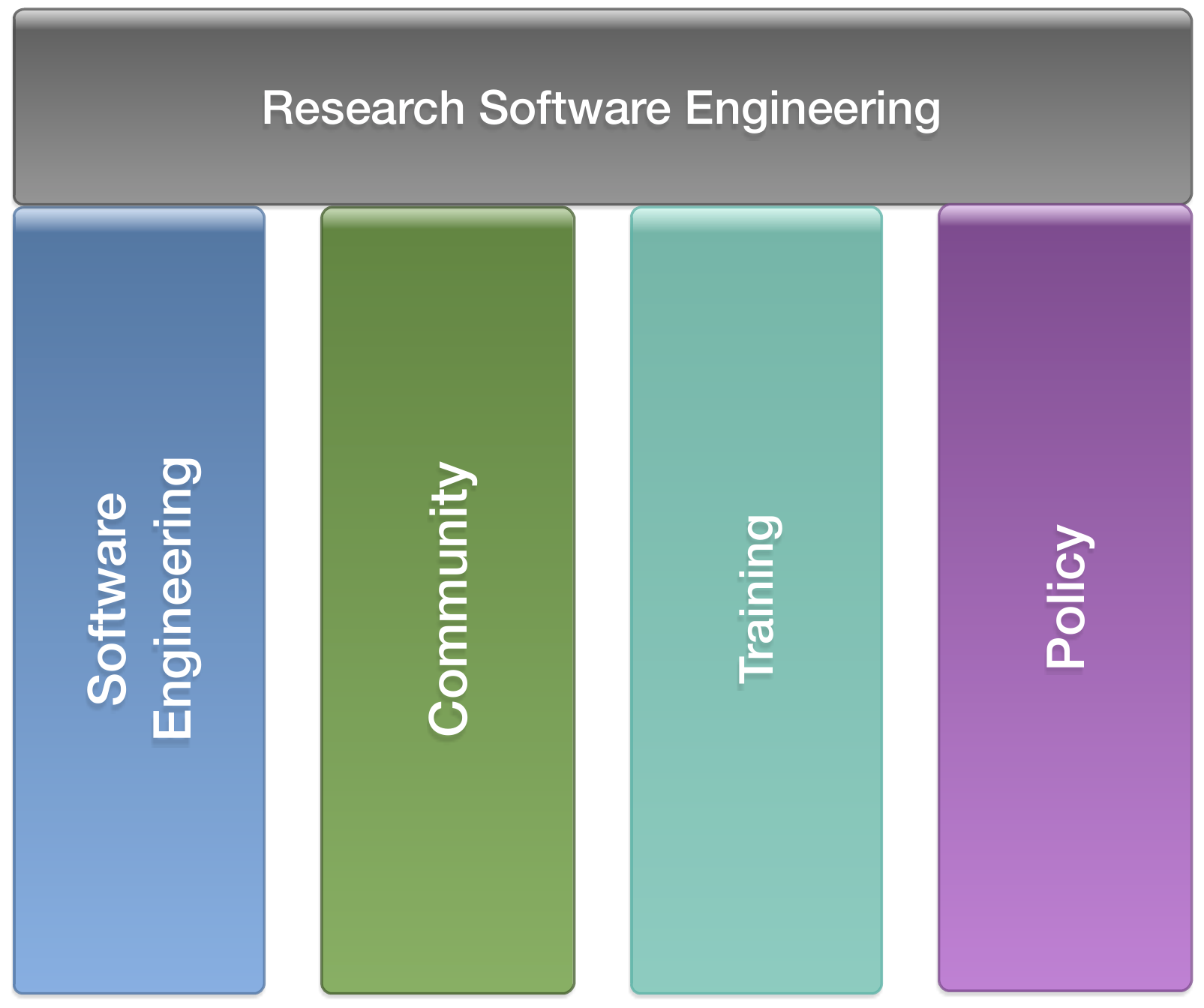}
\caption{The four supporting pillars of Research Software Engineering in our preliminary model.}
\label{figure:pillars}
\end{figure*}

\section{Core elements of Research Software Engineering}
\label{section:rs-activities}

Based on our observations of the way research software engineering has developed over the past few years, we see a series of activities that we believe contribute to an institution (or group of institutions) being able to provide an RSE offering to its research community that is sustainable and manageable over the long-term. These activities can be combined into groups covering specific areas which we define as the pillars of research software engineering---the key structures around which a successful research software development capability can be built. These pillars are:
\begin{itemize}
\item software engineering,
\item community,
\item training, and
\item policy.
\end{itemize}
Furthermore, it is our assertion that to be able to offer comprehensive research software support, activities from each of these pillars must be provided. 
We now provide an overview of the pillars, as highlighted in Figure~\ref{figure:pillars}:

\subsection{Software Engineering}

Software engineering encompasses the process of building software and the people who build it. The software aspects of research software development can involve any of a wide range of common languages, as well as sometimes including more obscure research-focused languages or Domain Specific Languages (DSLs). The process of building software in a research environment is, however, somewhat different to that which professional software engineers are likely to be familiar with, and can be incompatible with standard software development models or processes. For example, software in scientific research is generally developed to solve a specific research challenge meaning that it is often built without thought for its longer-term, wider use or maintenance, as highlighted in the work of Morris and Segal~\cite{ref:Morris09}. Building research software also generally requires a lot more interaction with the clients -- the researchers -- requiring the developer(s) to have a much greater understanding of the task being undertaken and to be able to work productively with the researcher or research team to understand what they are trying to achieve and take an active role in developing the approach used to find a solution. It is for reasons such as these that the people who build research software are so vital to this element of the model. While they have the expertise to build software, they also have an understanding of the research community, the research lifecycle and the process of working effectively with researchers. This gives research software developers a unique and valuable skill-set.

\subsection{Community}

Communities provide a forum through which RSEs can meet, network, and share ideas and technical knowledge. The value of this cannot be underestimated, especially when working as part of a small development team or research group, or working independently as the only developer within a project or group of non-computational researchers. Even when working in a larger team, the diversity of thoughts, ideas, and perspectives that comes from bringing together individuals from different disciplines and technical backgrounds can be enormously helpful in developing new ideas and approaches to solving technical problems.

Community is a separate entity from software development within our model but it should support the better development of software and provide a means to develop collaborations, and to access support and advice on technical questions or ideas. In turn, the software development aspects of research software engineering should support the community, providing technical challenges for discussion, interesting projects and use cases that can be presented in technical talks, and forums for meeting other community members who can offer advice and support to help all members of the community achieve the best they can from their work.

\subsection{Training}

Training is vital in ensuring the long-term maintenance and growth of skills and in keeping up to date with the latest developments in the research software domain. It also supports the development of skills amongst researchers who will provide the base for the next generation of RSEs. While change happens rapidly in all technical arenas, this is especially noticeable in research. New technologies are emerging at a substantial rate, and new libraries and tools emerging from the open source community can gain rapid adoption.

We consider training to be a pillar in its own right because it encompasses a number of activities that are separate from the community element of research software engineering. It also has clear links to both the software and community elements of the model. Training can exist within a community context but it can also be separate and attract a different group of people to those who form an RSE community. Training can cover the development of basic skills such as those provided through the Software and Data Carpentry movements~\cite{ref:software-carpentry, ref:data-carpentry}. It can also include specialist, domain-specific training such as that provided for High-Performance Computing topics through the PRACE community~\cite{ref:prace-training}. More specialised local training may be provided via a local RSE community and also through the running of technical seminars as part of a community's events programme, providing the link between the community and training pillars. The link to software development comes through the technical benefits and associated improvements in productivity, software quality, and robustness that can be expected by enhancing the skills of researchers and software developers.

\subsection{Policy}

Policy advances are also critical to enabling the broader system changes required to increase understanding of software's crucial role as enabling infrastructure for research, and to promote software as a principal component of cyberinfrastructure strategic planning. Cultural change in the research environment within which research software engineering work occurs is needed, at all levels, from departmental and institutional to international.

In addition to simply promoting research software engineering processes as valuable to researchers and Principal Investigators, there are a number of much more significant areas where achievement of more substantial change amongst institutions, funders, and the wider research community could offer important (and arguably essential) changes to the way that certain aspects of their processes are currently handled. These include: 

\begin{itemize}
\item Recognising software outputs as first-order research assets and providing means to assign credit to them
\item Providing better structured career pathways to help sustain research software engineering roles and to improve opportunities for career progression, alongside coherent approaches to training
\item Incorporating RSEs within funding guidelines
\item Providing researcher access to RSEs, which can include providing physical spaces for collaboration (see~\cite{ref:msdse-inst-change}).
\end{itemize}

These types of large-scale cultural changes can be brought about in a more structured manner through advocacy to key decision-makers by RSE community leaders, rather than in an ad hoc manner by individuals or small groups of developers.

Measurement is another key element of system change, providing evidence to decision-makers of the benefits of change, and analysing priority areas. The survey work on RSEs described above is already providing valuable confirmation of the key role of RSEs, and studies on the critical role of software in research add to the argument, such as the work of Nangia and Katz.~\cite{ref:Nangia17}. 

\section{A preliminary model for sustainable research software}
\label{section:initial-model}

This preliminary model is a work-in-progress and analysis is still ongoing to identify suitable solutions to a number of questions. The model described here is therefore intended to stimulate discussion and seek feedback with a view to developing a further paper presenting a more detailed, refined and more concrete description of the model. 

The basis for the model is the pillars described in Section~\ref{section:rs-activities}. It is believed that these pillars:
\begin{enumerate}
\item encapsulate the wealth of processes, topics and activities that make up research software engineering, and
\item represent, in their naming, the core, top-level concepts that individuals can identify with as being of utmost importance in enabling research software engineering.
\end{enumerate}

Other aspects that are required to complete the structure of the model are:

\begin{itemize}
\item Defining the processes that link or bring together the concepts represented by the pillars. This requires an understanding of the synergies between activities that fit within different pillars. For example, can links between training and communities offer greater combined benefits than just offering opportunities in one of the two areas?
\item Identifying the profiles of the individuals that each pillar relates to or targets, e.g. where do researchers, academics, software engineers, RSEs, research data specialists, research managers, etc. fit within the model? How significant is this in ensuring its success?
\item Identifying how generally applicable the current pillar definitions are. Do they apply differently in the context of individuals or groups? How robust are they in the light of possible structural changes in RSE communities? One possible way to look at this is in the context of the different levels of individual/team, research software and the wider field of research software that URSSI considers under their issues in the figure ``Key factors for URSSI conceptualization''~\cite{ref:urssi-newsletter}.
\end{itemize}

\subsection{Outstanding issues: questions and queries}

The points highlighted above need to be addressed in order to complete the initial structure of our model. However, this initial structure will then need refining. There are several more general questions that we feel will require further investigation/discussion as part of this model refinement process.

\begin{itemize}
\item Do the four pillars highlighted in Section~\ref{section:rs-activities} represent the complete picture? Are there any further pillars that should be defined? 

\item Are any of the topics covered by existing pillars of sufficient importance/significance that they should be promoted to form separate pillars in their own right?

\item Is the naming of the pillars correct? So far, these have been determined amongst a fairly small group of individuals with extensive experience in the RSE community. However, others, either from outside the RSE community or from different scientific backgrounds, may feel the pillars could be named differently, perhaps to clarify their meanings, or to provide a different slant on the way they are viewed from different perspectives.

\item Do groups or individuals relate differently to the pillars? Can we consider all individuals as being the same in the context of this question or are there differences in the way that individuals of different profiles identify with the pillars -- e.g. researchers or RSEs?

\end{itemize}

We hope to have the opportunity to investigate some of these issues with the wider research software community to gather thoughts, feedback and suggestions that can help to test, refine, and complete the preliminary model set out here.

\section{Conclusions and Future Work}
\label{section:conclusions}
\balance
Research software engineering has come a long way in the past six or so years. Nonetheless, it is still in its infancy as a discipline and while many different groups have emerged and different approaches have been, or are being, tested, we have observed that there is still a lack of significant formal structures or models that can be used to explain how and why RSE works in different contexts and, most importantly, how it can be effectively sustained and grown to offer its benefits to a much wider range of researchers. 

In this paper we have described work that is currently in progress to define a model that can offer one formalisation of a structure for research software engineering that brings together a full set of activities that we believe are necessary to provide a sustainable offering. Since the model is still in development, this paper seeks to gather feedback and thoughts on the perceived correctness of our proposed model and suggestions on how it might be improved. To this end, we have highlighted some of the specific questions that remain in developing the next iteration of the model and some more general points where we feel additional input is important.

Going forward, we intend to prepare a more detailed publication that defines our next iteration of the model, addressing the various issues raised here. As part of this ongoing work, we want to gather thoughts and feedback as part of a wider discussion on the ideas presented. There are two ways in which you can engage with this process: you can email the lead author, Jeremy Cohen, to express interest, or you can submit thoughts or questions to the \textit{rse-models} repository~\cite{ref:rse-models} that has been set up to capture and collect such information.

\section*{Acknowledgements}
JC acknowledges the support of the UK Engineering and Physical Sciences Research Council (EPSRC) through grant EP/R025460/1.

RH acknowledges the support of the University of Manchester for his UK and International RSE activities.

NCH acknowledges the support of the UK Engineering and Physical Sciences Research Council (EPSRC), Economic and Social Research Council (ESRC) and Biotechnology and Biological Sciences Research Council (BBSRC) through grant EP/N006410/1 for the Software Sustainability Institute.

\bibliographystyle{IEEEtran}
\bibliography{BuildSustainableStructureRSE}

\end{multicols}
\end{document}